\documentclass[11pt]{article}

\newcommand{\eref}[1]{(\ref{#1})}
\newcommand{\ket}[1]{\ensuremath {|\: #1 \: \rangle}}
\newcommand{\bra}[1]{\ensuremath{\langle \: #1 \:|}}
\newcommand{\braket}[2]{\ensuremath{\langle \: #1 \: | \: #2 \: \rangle}}

\newcommand{\ves}[2]{\ensuremath{#1_1,#1_2, \ldots, #1_{#2}}}

\newcommand{\llrr}[1]{\ensuremath{\left( #1\right)}}

\usepackage{graphicx}
\usepackage{subfigure}
\usepackage{amsmath}
\usepackage{algorithmic}

\usepackage{algorithm}
\renewcommand{\algorithmicrequire}{\textbf{Input:}}

\begin{document}

\title{An introduction to quantum annealing}


\author{Diego de Falco and Dario Tamascelli\\
\small{Dipartimento di Scienze dell'Informazione, Universit\`a degli Studi di Milano}\\
\small{Via Comelico, 39/41, 20135 Milano- Italy}\\
\small{e-mail:defalco@dsi.unimi.it,tamascelli@dsi.unimi.it}
}
\date{}
\maketitle

\begin{abstract}
Quantum Annealing, or Quantum Stochastic Optimization, is a classical randomized algorithm which provides good heuristics for the solution of hard optimization problems. The algorithm, suggested by the behaviour of  quantum systems, is an example of proficuous cross contamination between classical and quantum computer science. In this survey paper we illustrate how hard combinatorial problems are tackled by quantum computation and present some examples of the heuristics provided by Quantum Annealing. We also present preliminary results about the application of quantum dissipation (as an alternative to Imaginary Time Evolution) to the task of driving a quantum system toward its state of lowest energy.
%
\end{abstract}
%
%

%
%
\section{Introduction}
Quantum computation stems from a simple observation\cite{feyn82,deutsch85}: any computation is ultimately performed on  a physical device; to any input-output relation there must correspond a change in the state of the device. If the device is microscopic, then its evolution is ruled by the laws of quantum mechanics.\\[5pt]
The question of whether the change of evolution rules can lead to a breakthrough in computational complexity theory is still unanswered \cite{vazirani97,kempe06,laumann09}. What is known is that a quantum computer would outperform a classical one in specific tasks such as integer factorization \cite{shor97} and searching an unordered database \cite{grover96,grover01}. In fact, Grover's algorithm can search a keyword quadratically faster than any classical algorithm, whereas in the case of Shor's factorization the speedup is  exponential. However, since factorization is in NP\footnote{To be more precise, factorization is in FNP, the function problem extension of NP. The decision problem version of factorization (given an integer $N$ and an integer $M$ with $1 \leq M \leq N$, does $N$ have a factor $d$ with $1 < d < M$?) is in NP.}  but not believed to be NP-complete, the success of factorization does not extend to the whole NP class.\\
Nonetheless, quantum computation is nowadays an active research field and represents a fascinating example of two way communication between computer science and quantum physics. On one side, computer science used quantum mechanics to define a new computational model, on the other the  language of information and complexity theory has allowed a sharper understanding of some aspects of quantum mechanics.\\
Moreover, it is clear that quantum mechanics will sooner or later force its way into our classical computing devices: Intel$^{\mbox{\small{\textregistered}}} $ has recently (Feb 2010\footnote{http://www.intel.com/pressroom/archive/releases/2010/20100201comp.htm}) announced the first 25nm NAND logical gate: 500 Bohr radii.\\[5pt]
In this paper we present an introduction to quantum adiabatic computation and quantum annealing, or quantum stochastic optimization, representing, respectively, a quantum and a quantum-inspired optimization algorithm. In Section \ref{sec:transition}  we introduce the class of problems that are hard for a quantum computer. We consider the problem 3-SAT, its mapping into a quantum problem and the quantum adiabatic technique. In Section \ref{sec:QA} we describe Quantum Annealing, a classical algorithm which captures the features of the quantum ground state process to provide good heuristics for combinatorial problems. Section \ref{sec:dissipation} is devoted to dissipative quantum annealing and to the presentation of preliminary heuristic results on the dynamics of a dissipative system. Concluding remarks and possible lines of future research are presented in the last section.

\section{Transition to the quantum world} \label{sec:transition}
The need of considering the features of quantum computing devices led to the definition of the quantum counterpart of some classical complexity classes. Quantum algorithms are intrinsically probabilistic, since the result of the measurement of an observable of a quantum system is a random variable. Not surprisingly, the quantum correspondent of P is BQP (Bounded-error Quantum Polynomial-time), the class of decision problems solvable in polynomial time by a quantum Turing machine, with  error probability at most 1/3.\\[5pt]
QMA (Quantum Merlin Arthur) is the quantum counterpart of the class NP\cite{watrous00}. It is defined as the class of decision problems such that a ``yes'' answer can be verified by a 1-message quantum interactive proof. That is: a quantum state (the ``proof'') is given to a BQP (i.e. quantum polynomial-time) verifier. We require that if the answer to the decision problem is ``yes'' then there exists a state such that the verifier accepts with probability at least 2/3; if the answer is ``no'' then for \emph{all states} the verifier rejects with probability at least 2/3.\\[5pt]
All the known complete problems for the class QMA are promise problems, i.e. decision problems where the input is promised to belong to a subset of all possible inputs. An example of QMA-complete problem is the $k$-\emph{local Hamiltonian problem}:
\\[5pt]
INSTANCE: a collection $\{\ves{H}{n}\}$ of Hamiltonians each of which acts on at most $k$ qubits; real numbers $a,b$ such that $b-a = O(1/poly(n))$.\\[5pt]
QUESTION: is the smallest eigenvalue of $\sum_{j=1}^n H_j$ less than $a$ or greater than $b$, promised that this is the case?
\\[5pt]
It has been shown that this decision problem is complete for the class QMA even in the case of 2-local Hamiltonians\cite{kempe06}.\\
The well known problems $k$-SAT, which are complete for NP for $k\geq 3$, can be encoded in a $k$-local Hamiltonian problem.\\
For instance, let us consider  a given boolean formula:
\[
 C_1 \wedge C_2 \wedge \ldots \wedge C_M
\]
over the variables $x_1,x_2,\ldots,x_N$; each $C_i$ is the disjunction $\left ( b_{i_1} \vee b_{i_2}\vee \ldots \vee b_{i_k} \right)$ and $b_i$ is either $x_i$ or its negation $\neg x_i$. Our problem is to decide whether there is an assignment to the boolean variables $x_1,x_2,\ldots,x_N$ that satisfies all the clauses simultaneously.\\
Given a clause $C_i = \left ( b_{i_1} \vee b_{i_2}\vee \ldots \vee b_{i_k} \right)$ we define the local function:
\[
f_{C_i}(x_1,x_2,\ldots,x_N)  = \prod_{{i_j} \in \Lambda_+^i} (1-x_{i_j}) \cdot \prod_{{i_j} \in \Lambda_-^i} x_{i_j}
\]
$\Lambda_-^i,\ \Lambda_+^i$ containing all the indices of the variables that compare negated or not negated  in the clause $C_i$ respectively. Given an assignment $(x_1,x_2,\ldots,x_N)$, the function $f_{C_i}$ will return 0  if the clause $C_i$ is satisfied, 1 otherwise.\\
The cost $V(x_1,x_2,\ldots,x_N)$ of an assignment can be defined therefore in terms of the functions $f_{C_i}$ as:
\[
 V(x_1,x_2,\ldots,x_N) = \sum_{i=1}^M f_{C_i}(x_1,x_2,\ldots,x_N),
\]
that is the total number of violated clauses. An instance of $k$-SAT is therefore satisfiable if there exists a configuration of zero cost.
\\[5pt]
The translation of a $k$-SAT problem to the corresponding $k$-local Hamiltonian, problem is straightforward.\\
To each variable $x_i$ we assign a \emph{qubit}, i.e. a 2-level quantum system. For the sake of definiteness we will consider spins $\left (\sigma_1(i),\sigma_2(i),\sigma_3(i) \right )$, $ i=1,2,\ldots,N$ and their component $\sigma_3(i)$ along the $z$  axis of a given reference frame as computational direction. Furthermore, we decide that if an assignment assigns the value 1 ($true$) to the variable $x_i$ then the corresponding qubit is in the state $\sigma_3(i) = +1$ (spin up) and in the state $\sigma_3(i) = -1$ otherwise.\\
To each clause we associate the Hamiltonian term:
\begin{equation}\label{eq:Hi}
 H_i =  \prod_{{i_j} \in \Lambda_+^i} \frac{1-\sigma_3(i_j)}{2} \cdot \prod_{{i_j} \in \Lambda_-^i} \frac{1+\sigma_3(i_j)}{2},
\end{equation}
which acts non  trivially only on $k$ qubits. The total Hamiltonian:
\begin{equation} \label{eq:Htot}
 H = \sum_{i=1}^M H_i
\end{equation}
is $k$-local and plays the role of a cost function: if an assignment satisfies all the clauses simultaneously then the corresponding energy is zero. Otherwise the energy would be equal to the number of violated clauses, thus $>0$.
\\[5pt]
The $k$-local Hamiltonian problem associated to $k$-SAT is a particular case of the $k$-QSAT problem, which is formulated as follows:\\[5pt]
INSTANCE: Hamiltonian $H = \sum_{j=1}^M P_j$ acting on $N$ qubits, where each $P_j$ is \emph{projector} on a $2^k$ dimensional subspace of the whole ($2^N$) dimensional Hilbert space. 
\\[5pt]
QUESTION: Is the ground state energy $E_0$ of $H$ zero, promised that either $E_0=0$ or $E_0 > 1/poly(N)$?
\\[5pt]
If each $P_j$ projects on an element of the computational basis, we obtain the $k$-local Hamiltonian associated to $k$-SAT by the construction described above.
\\[5pt]
Interestingly enough it is proved that 4-QSAT is QMA complete whereas 2-QSAT is in P\cite{bravyi06}. For 3-QSAT the answer is not known.
\\[5pt]
Quantum Adiabatic Computation \cite{farhi01}  was welcomed by the quantum computing community due to the preliminary good results produced when applied to small random instances of NP-complete problems \cite{farhi01,farhi00,young08} and its equivalence to the quantum circuital model \cite{aha07}.\\[5pt]
The computational paradigm is based on the well known adiabatic theorem \cite{born28,amba04}. Given a time $T>0$ and two Hamiltonians $H_I$ and $H_T$ we consider the time dependent Hamiltonian:
\begin{equation} \label{eq:interpolazione}
 H(t)  =  t H_T + (T-t) H_I, \ 0 \leq t \leq T,
\end{equation}
or, equivalently,
\[
\tilde{H}(s) = H\llrr{\frac{t}{T}},\ 0 \leq s \leq 1.
\]
Let us indicate with \ket{s;e_k(s)} the \emph{instantaneous eigenvector} of $\tilde{H}(s)$ corresponding to the \emph{instantaneous eigenvalue} $e_k(s)$ with $e_0(s) \leq e_1(s) \leq \ldots \leq e_n(s)$, $ 0 \leq s \leq 1$ and by $\ket{\psi(s)}$ the solution of the Cauchy problem:
\begin{equation*}
\begin{cases}
 i \frac{d}{ds} \ket{\psi(s)} = \tilde{H}(s) \ket{\psi(s)}\\
 \ket{\psi(0)} = \ket{0;e_0(0)}.
\end{cases}
\end{equation*}
The adiabatic theorem states that, if the time $T$ satisfies
\begin{equation}\label{eq:Testimate}
 T \gg \frac{\xi}{g_{min}^2},
\end{equation}
where $g_{min}$ is the minimum gap
\[
 g_{min}= \underset{0 \leq s \leq 1}{\min} (e_1(s)-e_0(s))
\]
and
\[
\xi = \underset{0 \leq s \leq 1}{\max} \left | \bra{s;e_1(s)} \frac{d\tilde{H}}{ds} \ket{s;e_0(s)} \right |,
\]
then $ \left |\braket{1;e_0(1)}{\psi(1)} \right |$ can be made arbitrarily close to 1. In other words if we start in the state \ket{0;e_0(0)} we will end up in the ground state \ket{T;e_0(T)} of the target Hamiltonian $H_T$. In practical cases $\xi$ is not too large; thus the size of $T$ is governed by $g_{min}^{-2}$: the smaller $g_{min}$ the slower must be the change rate of the Hamiltonian if we want to avoid transitions (the so called Landau-Zener transitions \cite{zener34}) from the ground state to excited states.\\[5pt]
The adiabatic method can be used to find the unknown ground state of an Hamiltonian $H_T$. We can start from the known (or easy to prepare) ground state of an auxiliary Hamiltonian $H_I$ and consider  the time dependent convex combination \eref{eq:interpolazione}.\\
For example let us consider the 3-SAT problem\cite{farhi00}: we need to understand if the ground state of the Hamiltonian \eref{eq:Htot} has energy $0$ or not. We set the target Hamiltonian $H_T$ to $H$. As initial Hamiltonian we consider the Hamiltonian
\[
 H_I = -\sum_i \sigma_1(i),
\]
having a ground state easy to prepare (all the spins aligned along positive $x$ direction).\\
The time $T$ required to fulfill the hypotheses of the adiabatic theorem is then the cost in time of the algorithm and is determined by the gap $g_{min}$.\\[5pt]
The results obtained in the seminal paper by Farhi \emph{et al.} \cite{farhi00} on small random instances of 3-SAT suggested that the gap $g_{min}$ scaled polynomially with the problem size. However, subsequent results \cite{vanDam01,reich04,znidaric06,krovi09,amin09} proved that  $g_{min}$ can be exponentially small. It has been even shown that adiabatic quantum computation can perform worse than other heuristic classical and quantum algorithms on some instance of 3-SAT \cite{hogg03}.\\
Though these results do not prove that 3-SAT is QMA-complete, they hint nevertheless that 3-SAT is hard for quantum computers as well. Recent studies focused on determining \emph{why} 3-QSAT has instances  which are hard for adiabatic quantum computation and what general features they possess. The idea is to use typical, but not necessarily worst, random cases with a specific density of clauses $\alpha=\frac{M}{N}$ ($M$ number of clauses, $N$ number of variables) and to investigate how $g_{min}$ varies (on the average) as a function of $\alpha$. We refer the interested reader to \cite{laumann09,laumann09b,amin09} and references therein.
%
%
%
\section{Quantum annealing} \label{sec:QA}
\begin{figure}[!h]
 \centering
 \includegraphics[width=10cm]{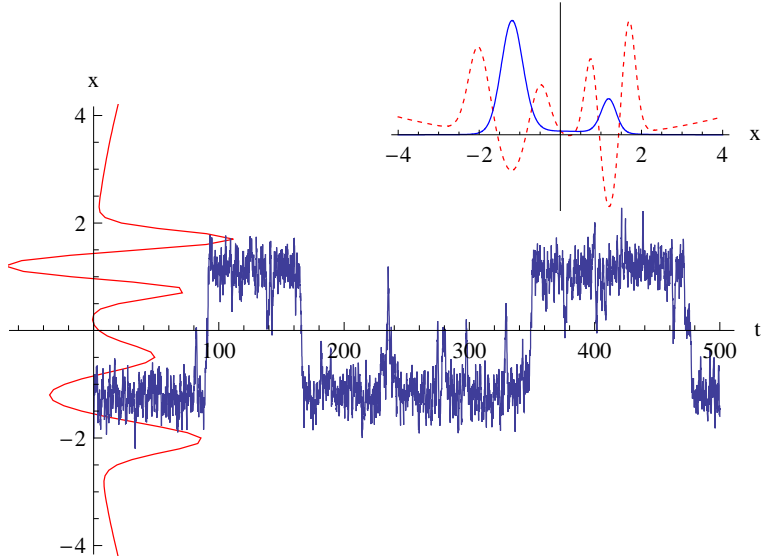}
\caption{A sample path of the ground state process $q_\nu$ for small $\nu$. The potential profile is reproduced on the vertical axes. In the inset we reproduce the potential $V(x)$ (dashed line) and the ground state  for the this example (solid line). For the example chosen the probability distribution is more concentrated around a local rather than the global minimum. What matters, however, is that the process jumps between stable configurations, the absolute minimum included.}\label{fig:figuratraj}
\end{figure}

\begin{figure}[!h]
 \centering
 \subfigure[]{\label{fig:finalqa}\includegraphics[width=5.4cm]{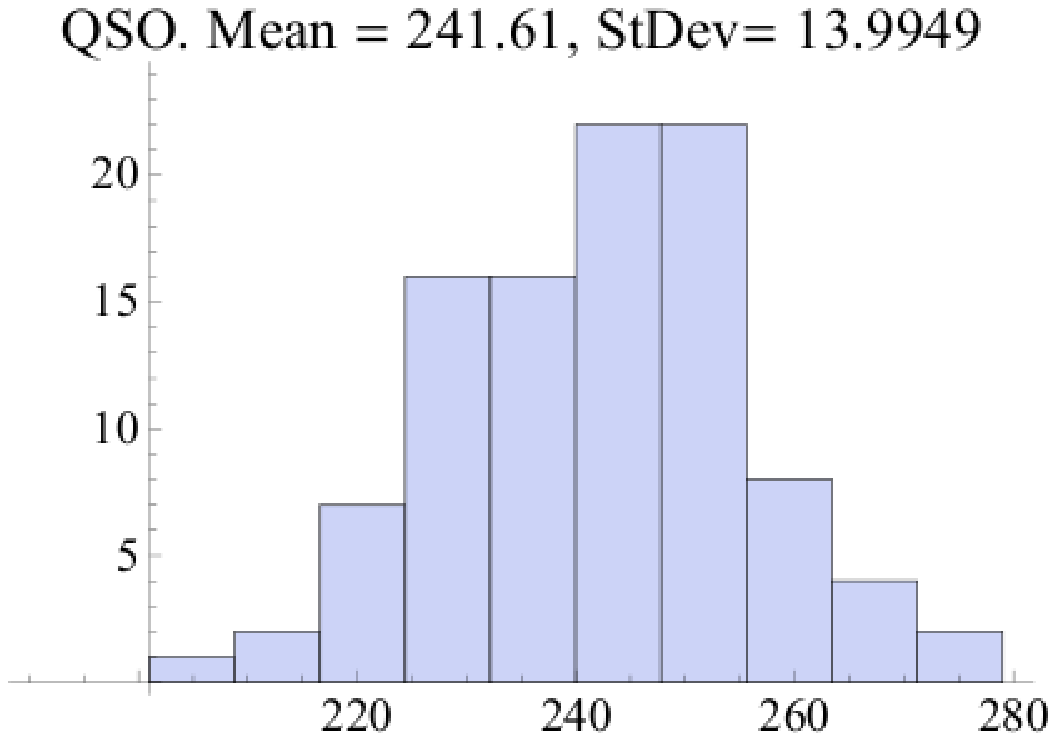}}
 \hspace{1cm}
  \subfigure[]{\label{fig:finalsa} \includegraphics[width=5.5cm]{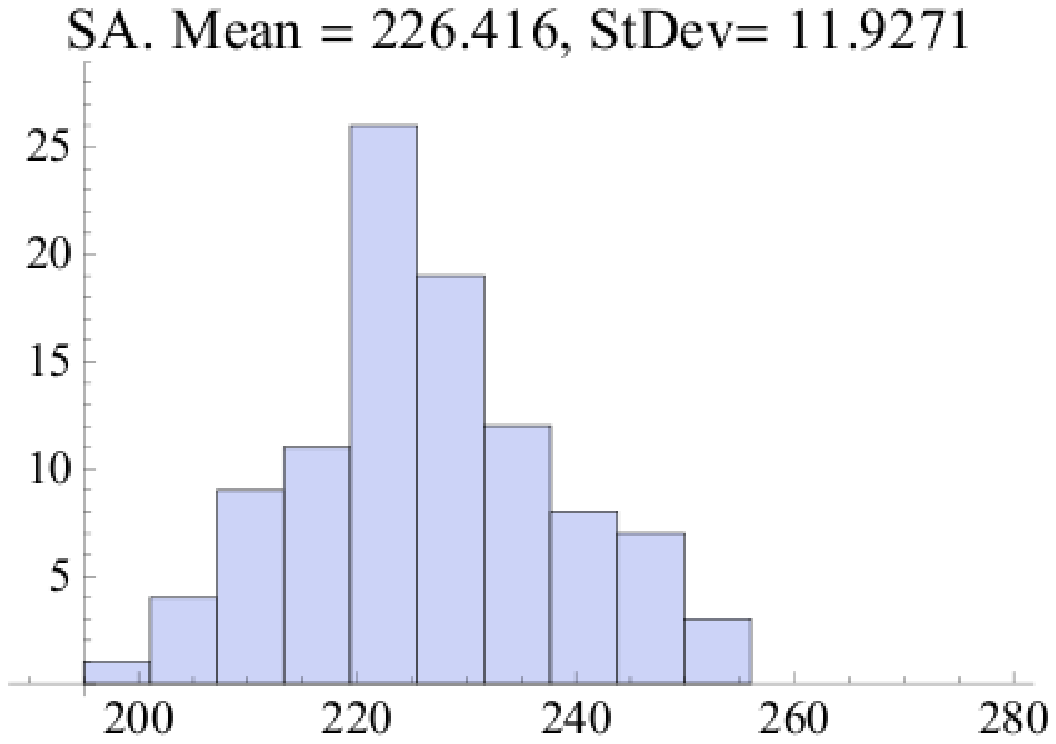}}
  \hspace{1cm}
  \subfigure[]{\label{fig:gqa}\includegraphics[width=5.4cm]{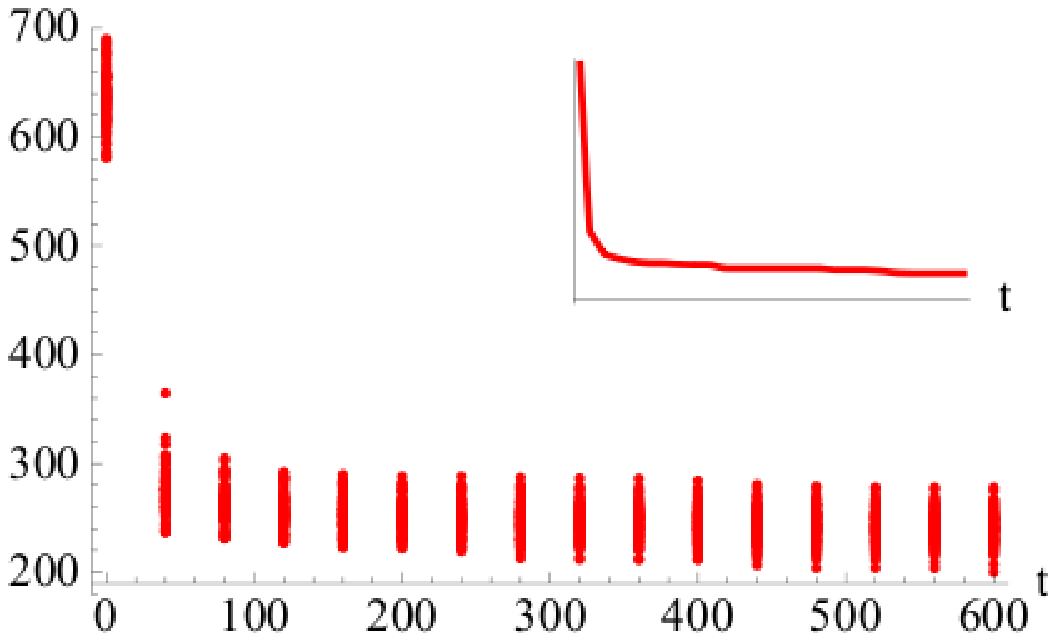}}
 \hspace{1cm}
  \subfigure[]{\label{fig:gsa} \includegraphics[width=5.5cm]{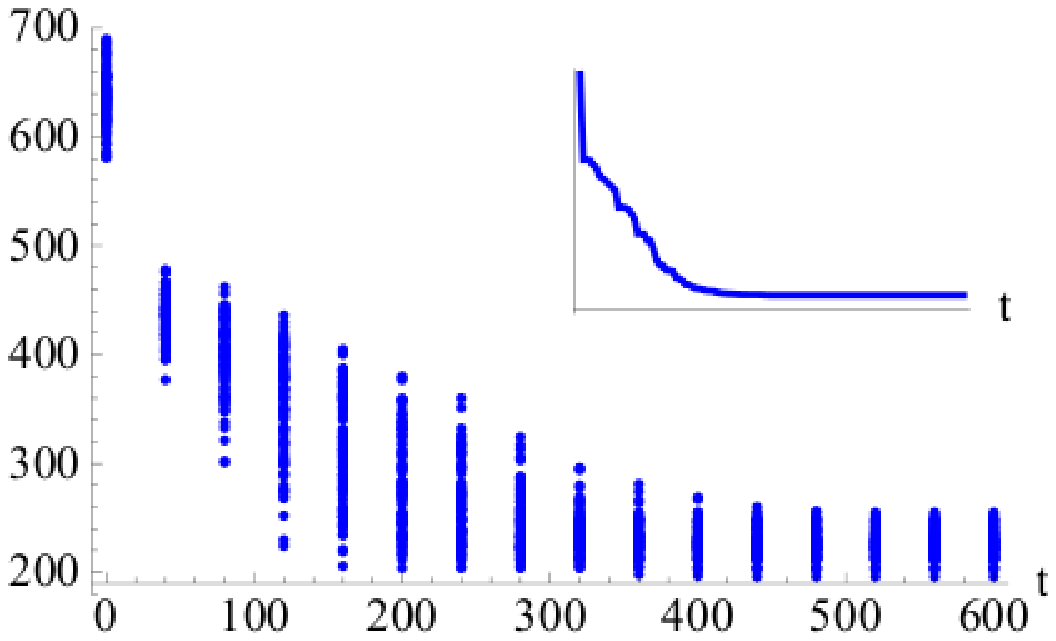}}
 \caption{Quantum Stochastic Optimization (QSO) vs. Simulated Annealing (SA). Benchmark: 100 random instances of $G_{500,0.01}$. Allocated CPU-time: 600 seconds. Parametrization: QSO: $\nu t=20$; $n=4$; $|Neigh|=5$; SA: geometric scheduling: temperature at $k$-th annealing step: $t_k= t_0(0.99)^k$;  $t_0=3.0$;  number of proposed moves per annealing step = $ 16 \cdot 500 $. (a) and (b) distribution of $V_{min}$ for QSO and SA respectively. (c) and (d) Values of $V_{min}$ reached respectively by QSO and SA on each instance of Graph Partitioning after every 50 (machine-time) seconds. In the inset of (c) and (d): values of $v_{min}$ along a single search path. }\label{fig:qasa}
\end{figure}

So far we discussed if and how quantum computational devices could be used to tackle problems which are hard to solve by classical means. In this section we present a class of classical heuristics (i.e. algorithms meant to be run on classical computational devices) suggested by the behaviour of quantum systems.
\\[5pt]
Many well-known heuristic optimization techniques \cite{papa98} are based on natural metaphors: genetic algorithm, particle swarm optimization, ant-colony algorithms, simulated annealing and taboo search. In simulated annealing \cite{kirk83}, for example, the space of admissible solutions to a given optimization problem is visited by a temperature dependent random walk. The cost function defines the potential energy profile of the solution space and thermal fluctuations avoid that the exploration gets stuck in a local minimum. An opportunely scheduled temperature lowering (annealing), then, stabilizes the walk around a, hopefully global, minimum of the potential profile.
\\[5pt]
The idea of using \emph{quantum}, instead of thermal, jumps to explore the solution space of a given optimization problem was proposed in Refs.\cite{defa88,apo89}\footnote{For the connection with the QAC of the previous section we refer the reader to \cite{santoro08}.}. It was suggested by the behaviour of the \emph{stochastic process} \cite{albe77,eleu94} $q_\nu$ associated with the ground state (state of minimal energy) of a Hamiltonian of the form:
\begin{equation} \label{eq:schr1}
H_\nu = -\frac{\nu^2}{2} \frac{\partial^2}{\partial x^2} + V(x),
\end{equation}
where the potential function $V$ encodes the cost  function to be minimized.
%
%
%
%
\begin{figure}[!h]
 \centering
 \includegraphics[width=10cm]{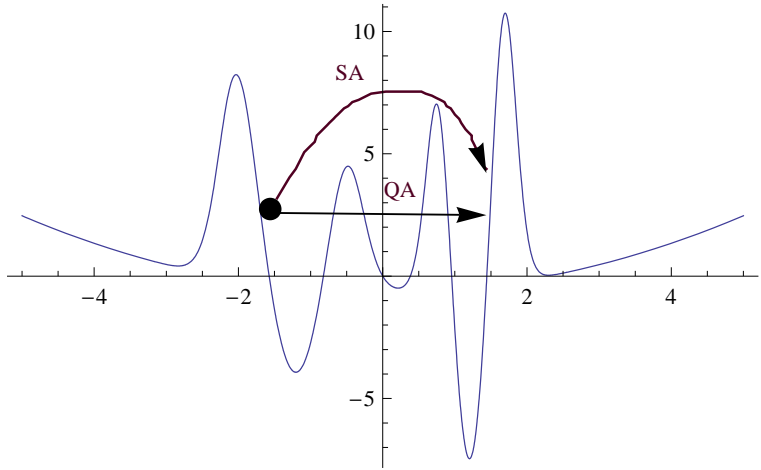}
\caption{Thermal jumps, which in SA allow the exploration of the solution space, are substituted, in QA, by quantum jumps (tunneling).}\label{fig:figura}
\end{figure}
\\For any fixed $\nu$, once given the ground state $\psi_\nu$ of $H_\nu$, the stochastic process $q_\nu$ can be built by the ground state transformation \cite{albe77}: let $\psi_\nu \in L^2(R,dx)$ be the ground state of the Hamiltonian \eref{eq:schr1}. Under quite general hypotheses on the potential $V$, the ground state $\psi_\nu$ can be taken strictly positive. The transformation:
\[
 U:\psi \rightarrow U\psi = \frac{\psi}{\psi_\nu},
\]
or \emph{ground state transformation}, is well defined and unitary from $L^2(R,dx)$ to $L^2(R,\psi_\nu^2 dx)$. Under this transformation, $H_\nu$ takes the form:
\[
 H_\nu = U H_\nu U^{-1} = -\nu L_\nu + E_\nu
\]
where
\[
 L_\nu = \frac{1}{2} \nu \frac{d^2}{dx^2}+ b_\nu
\]
has the form of the generator of a diffusion process $q_\nu$ on the real line, with drift
\[
 b_\nu(x) = \frac{1}{2} \frac{d}{dx} \ln\llrr{\psi_\nu^2(x)}.
\]
The behaviour of the sample paths of the stationary \emph{ground state process} $q_\nu$ is characterized by long sojourns around the stable configurations, i.e. minima of $V(x)$, interrupted by rare large fluctuations which carry $q_\nu$ from one minimum to another:  $q_\nu$ is thus allowed to ``tunnel'' away from local minima to the global minimum of $V(x)$ (see Fig.\ref{fig:figuratraj}). The diffusive behaviour of $q_\nu$ is determined by the Laplacian term in \eref{eq:schr1}, i.e. the kinetic energy, which is controlled by the parameter $\nu$. The deep analysis of the semi-classical limit performed in \cite{iona81} shows, indeed, that as $\nu \searrow 0$ \emph{``the process will behave much like a Markov chain whose state space is discrete and given by the stable configurations''}.
%
%
%
\\
Quantum annealing, or \emph{Quantum Stochastic Optimization} (QSO), in the original proposals of Refs.\cite{defa88,apo89}, did not intend to reproduce the dynamics of a quantum mechanical system (a task computationally untractable \cite{feyn82}), but rather to simulate the ground state process of the Hamiltonian $H_{\nu}$ as $\nu \searrow 0$. The algorithm is completely classical and intends to capture the features of the quantum process which can allow an efficient exploration of the solution space described above.
\\[5pt]
The desired ground state estimation is obtained by means of \emph{imaginary time evolution}, that is by letting an arbitrarily chosen normalized initial state $\psi(x,0)$ of the system evolve not under the action of  $e^{-itH_{\nu}}$ but under $e^{-tH_{\nu}}$. This replacement of the Schr\"odinger equation by a heat equation has the following property:
\[
\lim_{t \to \infty}\frac{1}{\alpha_t} e^{-t \frac{H_\nu}{\hbar}} \ket{\psi(0)} = \ket{\psi_\nu},
\]
where $\alpha_t = \braket{\phi_0}{\psi(0)} \exp(-tE_0)$. The proof of this fact is straightforward: given the eigenstates $\phi_0(=\psi_\nu), \phi_1,\ldots,\phi_N$ of $H_\nu$ and the corresponding eigenvalues $E_0 < E_1 < \ldots < E_N$, we have:
\begin{eqnarray}\label{eq:proiezione}
\lim_{t \to \infty} \frac{1}{\alpha_t}e^{-t \frac{H_\nu}{\hbar}}\ket{\psi(0)} &=& \lim_{t \to \infty}\frac{1}{\braket{\phi_0}{\psi(0)}}  \sum_{n = 0}^N \frac{e^{-t\frac{H_\nu}{\hbar}}}{e^{-t\frac{E_0}{\hbar}}}\ket{\phi_n}\braket{\phi_n}{\psi(0)} \\
&=& \lim_{t \to \infty}\frac{1}{\braket{\phi_0}{\psi(0)}}  \sum_{n = 0}^N e^{-t\frac{E_n - E_0}{\hbar}}\ket{\phi_n}\braket{\phi_n}{\psi(0)} \nonumber \\
&=& \lim_{t \to \infty} \ket{\phi_0} + \frac{1}{\braket{\phi_0}{\psi(0)}}  \sum_{n = 1}^N e^{-t\frac{E_n - E_0}{\hbar}}\ket{\phi_n}\braket{\phi_n}{\psi(0)}. \nonumber
\end{eqnarray}
The excited states are ``projected out'' and, asymptotically, only the ground state survives.\\
We are nevertheless left with the problem of determining $e^{- t H_\nu}$. A direct diagonalization of $H_\nu$ is impossible, given the dimensionality of the operator which scales exponentially with the dimension of the system. However, the Feynman-Kac formula \cite{Kac1949,Feynman1948}:
\begin{equation}
\label{eqn:feynmankac}
\left ( e^{-t H_\nu} \psi \right )(x) = E \left (\exp \left ( - \int_0^t{V(\varepsilon(\tau))\mathrm{d}\tau} \right )\psi(\varepsilon(t)) \middle | \varepsilon(0) = x \right ),
\end{equation}
suggests a way to \emph{estimate} the evolved state $e^{-t H_\nu}\psi$ for any initial state $\psi$: the component $x$ of the state vector corresponds to the expected value of (the exponential) of the integral of the potential (cost) function $V$ along the stochastic trajectories $\epsilon(\tau)$. Each trajectory $\epsilon(\tau)$ starts at $x$ and, in the time interval $(0,t)$, makes $N(t)$ transitions toward nearest-neighboburs with uniform probability; the number of random transition $N(t)$ is a Poisson process of intensity $\nu$. A detailed description of the sampling and of the ground state estimation process is beyond the scope of this paper and we refer the interested reader to Ref.\cite{apo89}. Here it suffices to say that once given an estimate of $\psi_\nu(y)$ for every neighbour $y$ of the current solution $x$, the exploration proceeds with high probability toward the nearest-neighbour solution having the highest estimated value of $\psi_\nu$. In Quantum Stochastic Optimization, therefore, each move is local (i.e. to a nearest-neighbour of the current solution) but the decision rule on which neighbour to accept is based on the prospection of an ensemble of long chains $\epsilon(\tau)$.
\\
When implemented in a working computer program, the procedure described above requires a variety of approximations. First of all, equation \eref{eqn:feynmankac} reproduces the ground-state only in the limit $t \to \infty$, corresponding to sample paths of infinite length. The expected value of the right hand side of \eref{eqn:feynmankac}, moreover, will be estimated by means of a finite size sample.  Finally,  the number of neighbours of a configuration may be too large to allow the estimation of $\psi_\nu$ on the whole neighbourhood of a given solution. The accuracy of the approximation depends, for example, on the actual length $\nu t$ of the sample paths, the number of paths $n$ per neighbour and the dimension  $|Neigh|$  of the subset of the set of neighbours of a given solution to consider at each move. Some ``engineering'' is also in order \cite{defa88} (see the pseudo-code of Quantum Stochastic Optimization reported here): for example, we can call a local optimization procedure every $t_{loc}$ quantum transitions. In addition, if the search looks to be stuck for in a local minimum, we can force a jump to another local minimum. \\
In Fig. \ref{fig:qasa} we show the results produced by QSO applied to random instances of the Graph Partitioning problem: given a graph $G=(V,E)$, where $V$ denotes the set of vertices and $E$ the set of edges, partition $V$ into two subsets such that the subsets have equal size and the number of edges with endpoints in different subsets is minimized. We extracted our instances from the family $G_{500,0.01}$ (i.e. random graph with 500 nodes and an edge between any two nodes with probability $0.01$.) used in Ref.\cite{johnson1989} and compare them with the results obtained on the same instances by Simulated Annealing. The comparison has been made by letting the programs implementing SA and QA run for essentially the same (machine) time. Simulated Annealing performs better than Quantum Stochastic Optimization: it finds, on the average, better (smaller mean value and variance) approximations of the best partition. What is interesting is that QSO goes down very fast toward a local minimum and then relies on quantum transitions to escape from it. In SA, on the contrary, a steep descend toward a local minimum could result in a early freezing of the search. We tested QSO also random satisfiable instances of 3-SAT with qualitatively similar results.
%
  \floatname{algorithm}{Procedure}
  \begin{algorithm}[p]
  \caption{Quantum annealing}
  \begin{algorithmic}  
    \REQUIRE initial condition $init$; control parameter $\nu$; duration $t_{max}$; tunnel time $t_{drill}$; local opt. time $t_{loc}$.
  \\[5pt]
   \STATE $t \leftarrow 0$;
   \STATE $\epsilon \leftarrow init$;
   \STATE $v_{min}=cost(\epsilon)$;
   \WHILE {$t < t_{max}$}
    \STATE $j \leftarrow 0$;
    \REPEAT
      \STATE $i \leftarrow 0$;
      \REPEAT
	\STATE $ \epsilon \leftarrow \mbox{Quantum Transition($\epsilon, \nu, t_{max}$)}$;
	\IF{$cost(\epsilon) <v_{min}$}
	  \STATE $v_{min} \leftarrow cost(\epsilon)$;
	  \STATE $i,j \leftarrow 0$;
	\ELSE
	  \STATE $i\leftarrow i+1$;
	\ENDIF
      \UNTIL{$i>t_{loc}$}
      \STATE $epsilon\leftarrow \mbox{Local Optimization($\epsilon$)}$.
      \IF {$cost(\epsilon)<v_{min}$}
	\STATE $v_{min} \leftarrow cost(\epsilon)$;
	\STATE $j \leftarrow 0$;
      \ENDIF
    \UNTIL {$j<t_{drill}$}
    \STATE draw a trajectory of length $\nu t_{max}$ and jump there.
    \STATE Local Optimization($\epsilon$)
   \ENDWHILE
   \end{algorithmic} 
   \end{algorithm}
  \floatname{algorithm}{Procedure}
  \begin{algorithm}[p]
  \caption{Quantum Transitions}
  \begin{algorithmic}
  \REQUIRE initial condition $\epsilon$; chain length $\nu t$; set of neighbours to estimate $Neigh$
  \\[5pt]
  \FORALL {neighbour  $ k \in\ Neigh$}
   \STATE estimate the wave function $\psi_\nu(k)$; 
  \ENDFOR
   \STATE $best \leftarrow \mbox{select a neighbour in $Neigh$ with probability proportional to $\psi_\nu$}$
   \RETURN $best$
   \end{algorithmic} 
   \end{algorithm}
  \floatname{algorithm}{Procedure}
  \begin{algorithm}[p]
  \caption{Local optimization}
  \begin{algorithmic}
  \REQUIRE initial condition $\epsilon$.
  \\[5pt]
  \RETURN the best solution found by any steepest descent strategy.
  \end{algorithmic} 
   \end{algorithm}
%
%
\\[5pt]
Recent variants of QSO,  as Imaginary Time Quantum Monte Carlo (ITQMC) \cite{car05,santoro06,santoro08,kuri09}, introduce a proper annealing schedule in the algorithm: the control parameter $\nu(t)$ is suitably reduced during the algorithm, as the temperature is in  reduced in SA.  Between each ``annealing'' step, the ground state $\psi_\nu$ is estimated by the same estimation procedure used in QSO. For some optimization problems, such as the Graph partitioning described above, ITQMC performs better than SA, for others,  e.g. 3-SAT, it does not. There is therefore no a-priori guarantee that QA will produce better heuristics than SA \cite{battaglia05}.
\\[5pt]
Interestingly enough, it has been shown that ITQMC performs at least as well as the \emph{real time} adiabatic approximation \cite{nishi08} described in the previous section. The intuitive reason behind this result has been clearly stated in \cite{amara93,santoro05}: the projection mechanism proper of imaginary time makes the evolution of the initial condition much more stable than the one determined by the Schr\"odinger equation. The adiabatic theorem, in fact, relates the energy gap between the ground state and the first excited state to the annealing time: the smaller the energy gap, the slower must be the change of the Hamiltonian $H_A(t)$ in order to avoid Landau-Zener transitions of the system from the ground to excited states (which do not correspond to the solution we are looking for). Since the projection mechanism suppresses the amplitude associated to excited states by accidental Landau-Zener transitions, the interpolation between the initial Hamiltonian $H_A(0)$ and the target Hamiltonian $H_T$ (the annealing) can be carried out more rapidly.
\section{Dissipative dynamics} \label{sec:dissipation}
\begin{figure}[htp]
 \centering
 \includegraphics[width=5.5cm]{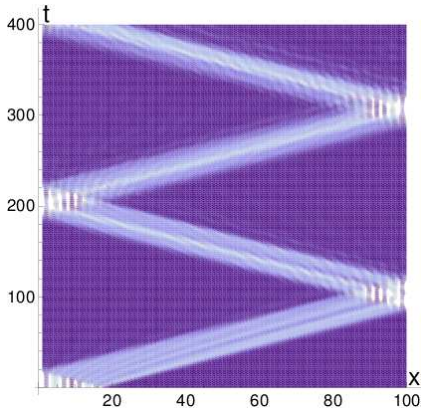}
 \caption{\label{fig:ballistic} $s=100, \lambda = 1$. The excitation, initially localized at the beginning of the chain, moves ballistically forth and back.}
 \end{figure}

\begin{figure}[htp]
 \centering
 \subfigure[]{\label{fig:bloch}\includegraphics[width=5.4cm]{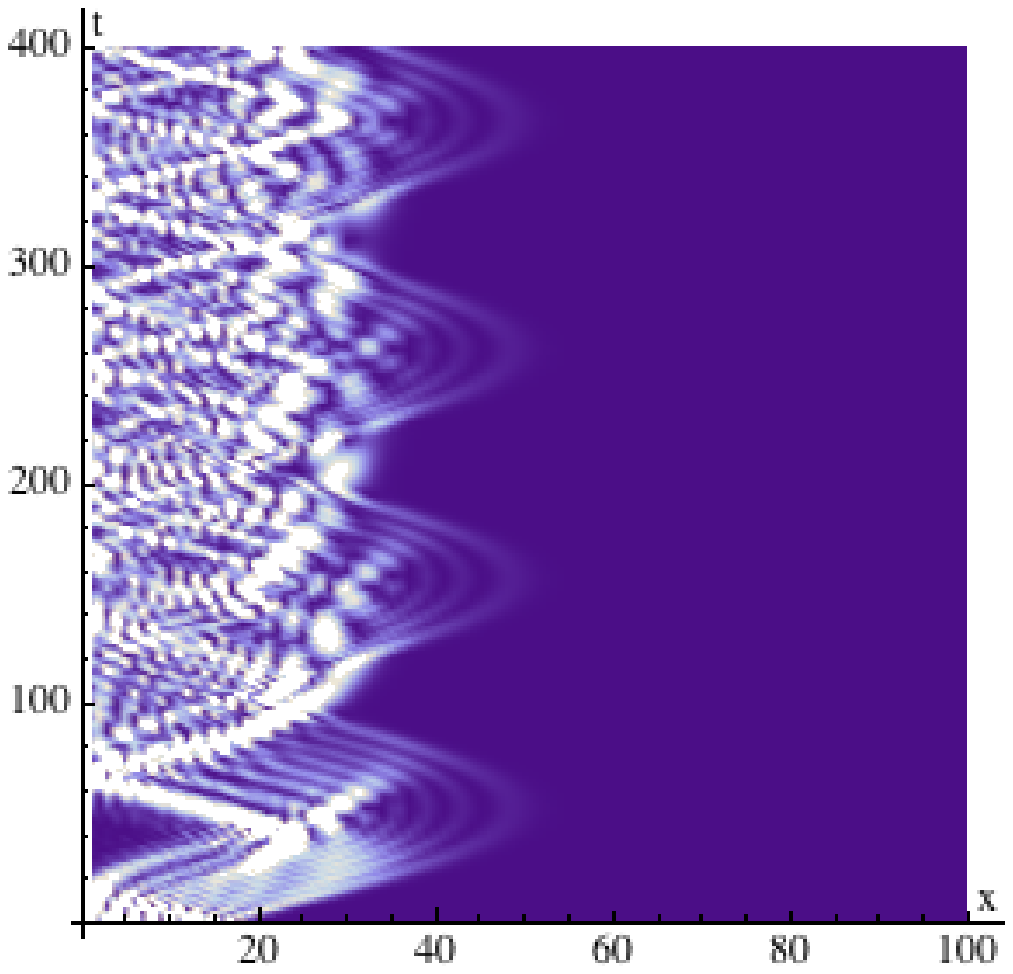}}
 \hspace{1cm}
  \subfigure[]{\label{fig:nobloch} \includegraphics[width=5.5cm]{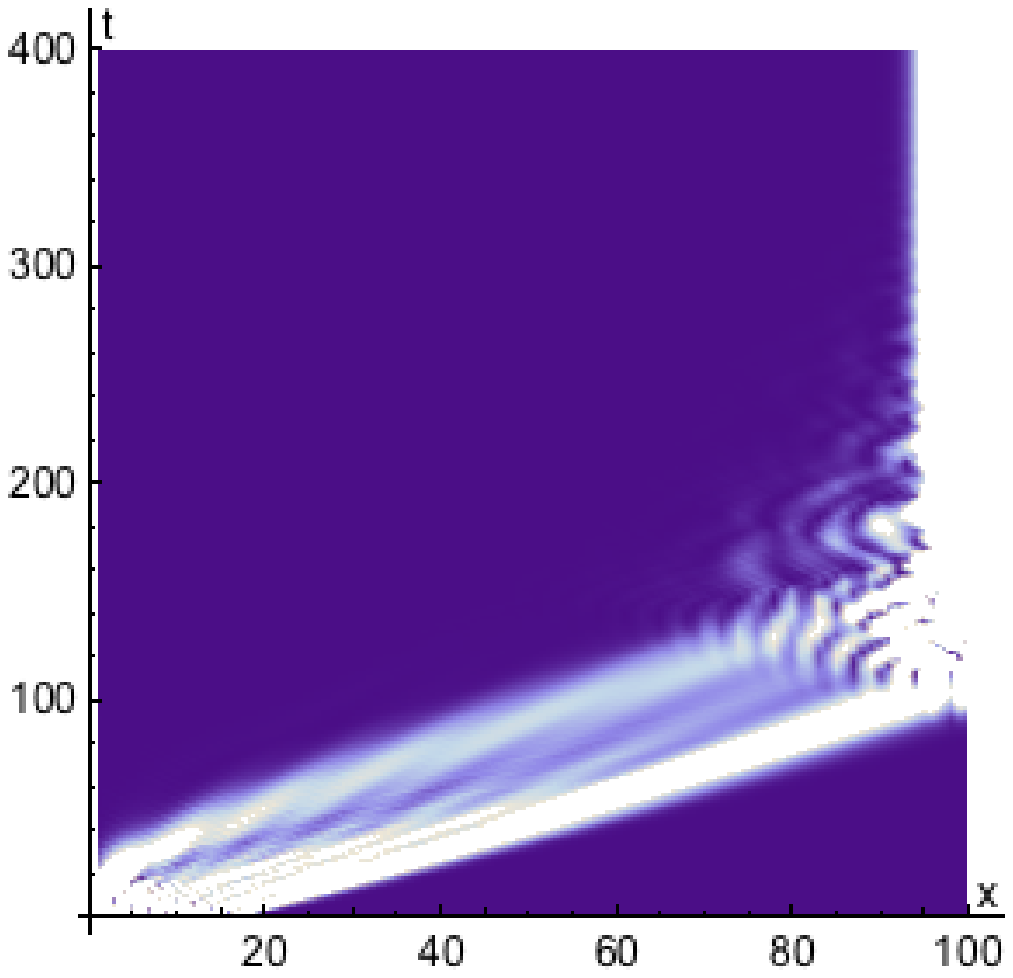}}
 \caption{$s=100,\ 0\leq t \leq 20s$. Frame (a): $g = 6/s,\ \beta=0$: the wave packet gets confined due to Bloch reflection. (b): for $g = 6/s,\ \beta=8/s$: Bloch reflections are suppressed by friction.}
\end{figure}

\begin{figure}[htp]
 \centering
 \subfigure[]{\label{fig:anderson}\includegraphics[width=5.5cm]{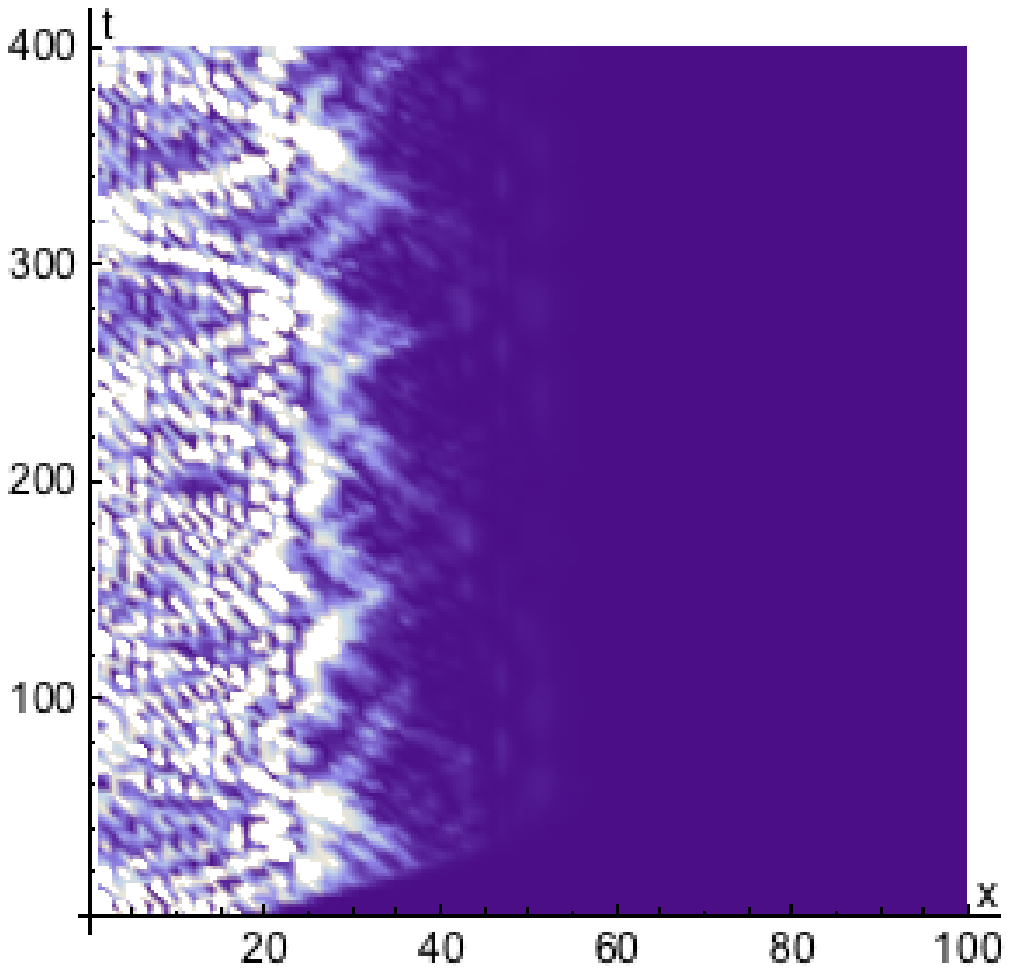}}
 \hspace{1cm}
  \subfigure[]{\label{fig:noanderson}\includegraphics[width=5.5cm]{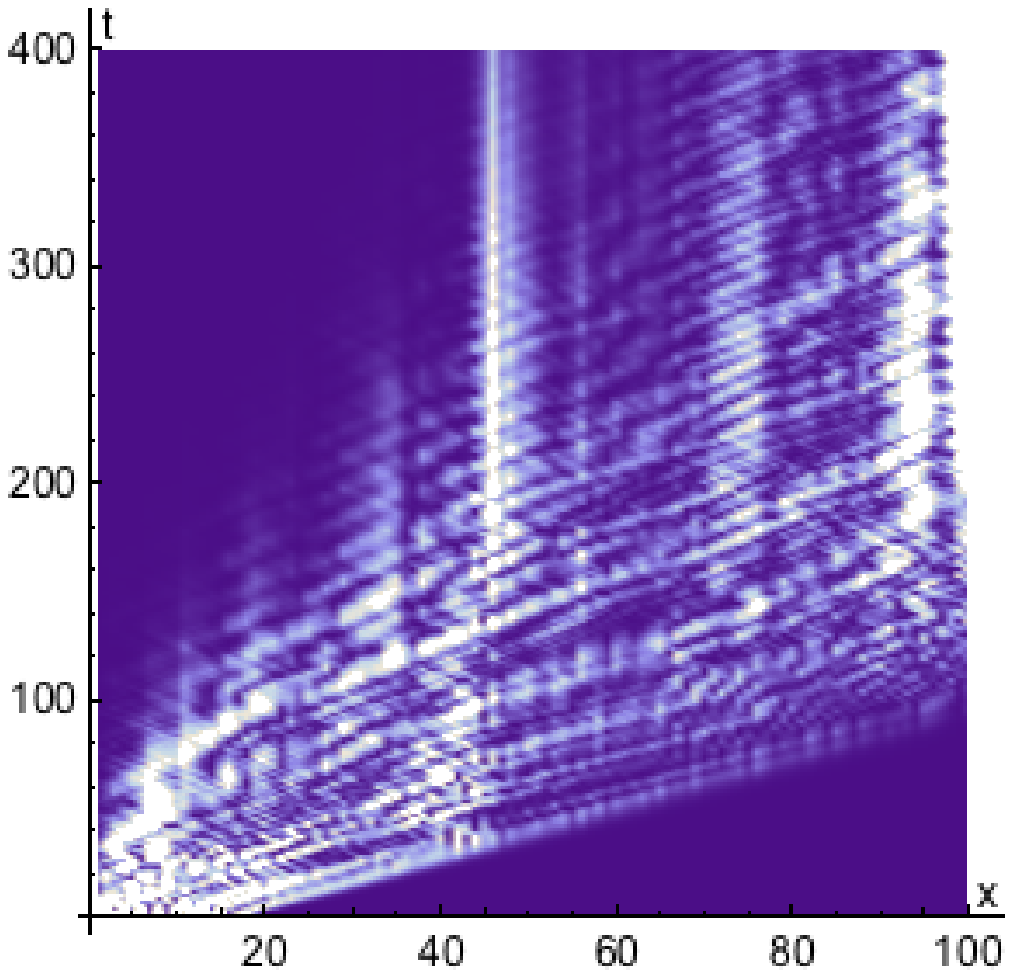}}
 \caption{\label{fig:fanderson} $s=100,\ 0\leq t \leq 20s$. Frame (a): $g = 0,\ \beta=0$: The wavepacket is localized due to Anderson localization  (b): for $g = 6/s,\ \beta=8/s$: friction suppresses Anderson localization: the probability mass percolates through the imperfections (additional random potential extracted from a normal population of mean 0 and standard deviation $\sigma = 0.06$) of the spin chain.}
\end{figure}

It becomes quite natural to ask whether a ``projection'' mechanism similar to the one operated by the  unphysical imaginary time evolution is available in some real time quantum dynamics. If it were possible, we could hope to exploit it to speed up quantum adiabatic computation without worrying too much about Landau-Zener transitions. \emph{Dissipative quantum annealing}, a novel model proposed in  \cite{defa09,tama09}, represents one first step in answering this question.\\
Given the usual Hamiltonian $H_\nu$, we add a \emph{non-linear} term, the Kostin \emph{friction} \cite{kostin72,kostin75}, modeling the effective interaction of the quantum system with an environment which absorbs energy. In the continuous case the Schr\"odinger-Kostin equation reads:
\begin{equation} \label{eq:kostin}
i \frac{d}{d t} \psi =  H_\nu \psi + \beta K(\psi),
\end{equation}
where $H_\nu$ is the Hamiltonian of \eref{eq:schr1} and
\begin{equation}
  K (\psi) = \frac{1}{2i} \log \llrr{\frac{\psi}{\psi^*}}.
\end{equation}
By rewriting the state $\psi(x)$ \emph{\`a la} de Broglie, 
\begin{equation}\label{eq:debroglie}
\psi(x) = \sqrt{\rho(x)}\  e^{iS(x)},
\end{equation}
the nonlinear part $K$ of the Hamiltonian \eref{eq:kostin} assumes the form $K(\sqrt{\rho(x)}\ e^{iS(x)}) = S(x)$, whose gradient corresponds to the current velocity \cite{Messiah}. This justifies the names \emph{friction} for the term $K$ and \emph{friction constant} for $\beta>0$.\\
Given a solution $\psi(t)$ of the Schr\"odinger equation $i \frac{d}{dt} \psi = (H_\nu + K)\psi$ the following inequality holds:
\[
 \frac{d}{dt} \bra{\psi(t)}H_\nu \ket{\psi(t)} \leq 0.
\]
It means that the energy of the system is a monotone non increasing function of time: the system dissipates energy.\footnote{We point out that the full interaction with the environment would include a random force describing the back action of the environment onto the system. We are then well aware of the fact that we are considering just the dissipative part of a fluctuating-dissipating system. Indeed, the complete Kostin equation was the correspondent in the Schr\"odinger representation of the Quantum Langevin Equation, that is the equation for the observables of an harmonic oscillator coupled to a bath of oscillators \cite{ford65} in the thermodynamic limit. The proper language to describe such an interaction would be the theory of open quantum systems \cite{breuer2002}. We decided, however, to adopt a phenomenological approach such as the one used in \cite{griffin76} and more recently, in the context of time dependent density functional theory,  in \cite{aspuru09}. A proper analysis of the complete system-environment interaction is currently under study.}\\[5pt]
In Ref.\cite{defa09} we showed that friction can play a useful role in suppressing two genuinely quantum effects which can affect the exploration of the solution space: Bloch oscillations  and Anderson localization.
Bloch oscillations \cite{bloch29} are due to the relation $v=\sin{p}$ between momentum $p$ and velocity $v$  of a particle moving on a regular lattice: as the momentum of the particle increases monotonically, the velocity can change sing and the particle starts moving backward along the lattice.\\
Anderson localization \cite{anderson58}, instead, appears when the lattice presents irregularities, manifesting themselves as a random potential on the sites of the lattice. The state of a quantum particle moving in a highly irregular potential landscape is spatially localized and the probability of tunneling through large regions is exponentially suppressed.\\
In \cite{defa09} we considered the simple case of a spin chain of finite size $s$ governed by an $XY$ Hamiltonian, that is:
\begin{equation}\label{eq:freeham}
 H_{XY} = \sum_{j=1}^{s-1} \lambda \llrr{\sigma_1(j) \sigma_1(j+1) + \sigma_2(j) \sigma_2(j+1) }.
\end{equation}
We took an initial configuration having a single spin up and all the others down. The spin up plays the role of a \emph{quantum walker} and, by a suitable choice of the initial conditions \cite{defa06a}, it behaves as an excitation moving ballistically along the chain as in Figure \ref{fig:ballistic}.
We introduce a simple potential/cost function $V(x) = -g x$. The motion of the walker is affected as shown in Figure \ref{fig:bloch}: Bloch oscillations appear hindering an exhaustive exploration of the solution space. As Figure \ref{fig:nobloch} shows, the oscillations can be suppressed by adding the frictional term, in the discretized form:
\begin{equation}
 (K\psi) (x)= \sum_{y=2}^x  \sin(S(y)- S(y-1)),
\end{equation}
to the system Hamiltonian. Here the phase function $S$ is defined as in Eq.\eref{eq:debroglie}.\\
Moreover, the convergence toward the ground state is witnessed by a progressive concentration of the probability mass at the end of the chain were the ground state is localized.\\[5pt]
Potential profiles determined by optimization problems are usually quite irregular. Figures \ref{fig:anderson} shows the consequence of the addition of a random potential to the free Hamiltonian \eref{eq:freeham}: the probability amplitude remains trapped in the first half of the chain, because of Anderson localization. Figure \ref{fig:noanderson} shows how dissipation can contrast Anderson localization: the probability mass percolates through irregularities and concentrates in the region where we chose to localize the ground state by adding a linear potential $V(x) = -gx$ as in the previous example. 
\\
\section{Conclusion and outlook}
Quantum Annealing is an example of cross contamination between two different research areas: computer science and physics. The ``constructive interference'' goes in two ways: on one side, a genuinely quantum effect, \emph{tunneling}, suggested a completely classical approximation algorithm which provides good heuristics to hard combinatorial problems. In this paper we presented in details the earliest version of Quantum Annealing, Quantum Stochastic Optimization.\\
On the other side, a trick used to estimate the state of minimal energy of a quantum system by classical means, Imaginary Time evolution \eref{eq:proiezione}, suggested the idea of exploiting dissipation to stabilize the evolution of quantum systems.\\
A mechanism able to suppress Anderson localization would be most welcome in Quantum Adiabatic Computation. In fact, Altshuler \emph{et al.} in Ref.\cite{krovi09} advanced the conjecture that \emph{``Anderson localization casts clouds over adiabatic quantum optimization''}.
The preliminary results presented here on the effects Kostin friction on simple toy models are promising: Anderson localization is suppressed. Still, whether an analogous mechanism can be of some use when solving hard optimization problems, together with a quantitative assessment of the results presented in \cite{tama09}, is an open research problem.

\end{document}
...................